\documentclass[hidelinks,runningheads]{llncs}
\usepackage{graphicx}
\usepackage{hyperref}
\usepackage[misc,geometry]{ifsym}
\begin{document}
\title{How a 4-day Work Week affects Agile Software Development Teams}
%
%
\author{Julia Topp \Letter\inst{1} \and
Jan Hendrik Hille \inst{1} \and 
Michael Neumann \inst{1} \and
David Mötefindt \inst{2}
}

\authorrunning{J. Topp et al.}
%
\institute{Hochschule Hannover - University of Applied Sciences and Arts, Ricklinger Stadtweg 120, 30459 Hannover, Germany\\
\{julia.topp,jan-hendrik.hille\}@stud.hs-hannover.de, michael.neumann@hs-hannover.de
\and
Agile Move, Ackerstr. 16, 30851 Langenhagen, Germany\\
info@agile-move.de
}
\maketitle              
\begin{abstract}
\textit{Context:} Agile software development (ASD) sets social aspects like communication and collaboration in focus. Thus, one may assume that the specific work organization of companies impacts the work of ASD teams. A major change in work organization is the switch to a 4-day work week, which some companies investigated in experiments. Also, recent studies show that ASD teams are affected by the switch to remote work since the Covid 19 pandemic outbreak in 2020. \textit{Objective:} Our study presents empirical findings on the effects on ASD teams operating remote in a 4-day work week organization. \textit{Method:} We performed a qualitative single case study and conducted seven semi-structured interviews, observed 14 agile practices and screened eight project documents and protocols of agile practices. \textit{Results:} We found, that the teams adapted the agile method in use due to the change to a 4-day work week environment and the switch to remote work. The productivity of the two ASD teams did not decrease. Although the stress level of the ASD team member increased due to the 4-day work week, we found that the job satisfaction of the individual ASD team members is affected positively. Finally, we point to affects on social facets of the ASD teams. \textit{Contribution:} The research community benefits from our results as the current state of research dealing with the effects of a 4-day work week on ASD teams is limited. Also, our findings provide several practical implications for ASD teams working remote in a 4-day work week. 

\end{abstract}
\section{Introduction}
In the last two decades, agile approaches became state-of-the-art in the area of software development \cite{VersionOne.2021}. The agile manifesto was developed 20 years ago in order to provide a common understanding of values and principles \cite{Beck.2001}. Agile methods are characterized by an intensified involvement of stakeholders and many interactions among the team members \cite{Abrahamsson.2002}. This focus on collaboration and communication is manifested by agile practices \cite{Williams.2010}, which are described in the guidelines of well-known agile methods like Scrum \cite{Schwaber.2020} or Extreme Programming~\cite{Beck.2000}.

Alternative work schedules and the 4-day work week in particular is a topic of interest in research and practice, which goes back to the 1970s \cite{Goodale.1975}. Though the concept of a 4-day work week is not established in practice, several companies \cite{Chappel.2019} and public administrations \cite{Wadsworth.2016} implemented experiments and pilots to test the effects of a compressed work week. Several studies present empirical findings concerning the effects of a compressed work week (e.g.,\cite{Chow.2006,Dunham.1987}). For instance, Facer and Wadsworth investigate the effects of a compressed work week schedule on facets like the employee satisfaction and the work-life balance \cite{Facer.2008}. They emphasize that the productivity of the employees is positively influenced. Further, they did not find a significant change concerning job satisfaction or work-life balance. One can assume, that a switch to a compressed 4-day work week may affects the work of agile software development teams due to the high relevance of social aspects of agile methods, as described above. 

Another facet related to alternative work forms and work organization types can be observed by the effects of the Covid 19 pandemic and the switch to remote work \cite{Nolan.2021}. As remote work seems to be a suitable solution to keep companies in business and their employees safe, many companies worldwide sent their employees to work remotely from home. Several studies describe the switch to remote work as a challenge for agile software development teams, as it affects teams collaboration, communication, productivity (e.g., \cite{Butt.2021,Ralph.2020,Schmidtner.2021}) and performance (e.g., \cite{Marek.2021,Neumann.2021,Oconnor.2021}). The switch to remote work also lead to a adaption of agile practices and roles in use \cite{DaCamara.2020}. For instance, the methodological implementation of agile practices like estimation techniques or retrospective and review meetings are affected by the virtualization of the collaboration \cite{Neumann.2021,Oconnor.2021}. Schmidtner et al. emphasize the effects on future work in agile software development \cite{Schmidtner.2021}. They point to the expectation of agile software development team members and experts that the remote work and use of tools will increase. 

In this study, related questions concerning a 4-day work week and the switch to remote work are addressed to teams of a global company using agile methods in their software development departments. A pilot of a 4-day work week was introduced at the beginning of 2021. Our key objective of the study is the investigation and analysis of the effects on agile software development teams working in a 4-day work week and a remote working environment. Thus, we defined the following research questions: 

\textbf{RQ 1:} Does the 4-day work week affect the use of agile methods? If so, how do teams adapt to the new circumstances?

\textbf{RQ 2:} Do the 4-day work week and remote work affect the productivity of agile software development teams?

\textbf{RQ 3:} Do the 4-day work week and remote work affect the job satisfaction and stress level of agile software development team members?

\textbf{RQ 4:} How does the 4-day work week combined with remote work due to the Covid 19 pandemic affect the social culture of agile software development teams?

This paper is structured as follows: In Section \ref{S2_RelWork}, we present the related work. We explain the selected research design in Section \ref{S3_RDesign}. We present the results of the paper at hand in Section \ref{S4_Results} and discuss our findings based on the research questions in the subsections \ref{S4-2_RQ1}, \ref{S4-2_RQ2}, \ref{S4-2_RQ3} and \ref{S4-2_RQ4}. We describe the Threads to Validity in Section \ref{S5_ThreatstoValidity} before the paper closes with a summary in Section \ref{S6_Conclusion}.

\section{Related Work}
\label{S2_RelWork}
In order to identify related work, we searched for studies and surveys, which are dealing with topics close to our context. In this section, we present an overview of the related work. We start with the recent studies dealing with the effects of the remote work during Covid 19 and close the section with the related work on 4-day work week.

The effects of the switch to remote work before the Covid 19 pandemic on agile software development teams have been barely investigated in recent years. However, against the backdrop of the Covid 19 pandemic, the topic gained in importance. As a result, several studies have been published dealing with the influences of the switch to remote work during the pandemic, the accompanying changes in agile software development team work organization, and the challenges of the new circumstances \cite{Nolan.2021}. 

Various studies dealing with the influence on productivity and performance of agile software development teams during Covid 19. However, the results presented in the recent studies show differences. Butt et al. investigated the positive and negative effects on agile software development teams during the pandemic in early 2021 by setting productivity in focus~\cite{Butt.2021}. The authors found that the productivity of agile software development teams decreased due to a minor coordination in the teams. Russo et al.~\cite{Russo.2021a} present in their study a correlation between the well-being and productivity. They point to the increased well-being of team members during the pandemic. In contract, Ralph et al., which are also dealing with the correlation of well-being and productivity, found that the productivity of agile software development teams decreased during the Covid 19 pandemic \cite{Ralph.2020}. The finding of a decreased productivity is also presented by Schmidtner et al. \cite{Schmidtner.2021}. 

Neumann et al. investigated the effects on the performance of agile software development teams during the Covid 19 pandemic \cite{Neumann.2021}. They found that the perceived performance of German agile software development teams did not decrease due to the switch to remote work.  The authors emphasize the positive influence of an increased transparency of the development process and the agile artifacts in use. Another qualitative study presented by O Connor et al. shows a positive effect on the performance of agile software development teams \cite{Oconnor.2021}. Furthermore, Marek et al. do not find significant changes concerning the performance of agile software development teams in their survey results \cite{Marek.2021}. 

Various studies show, that agile software development teams are able to rapidly react to the new circumstances due to the switch to remote work. The adaptions mainly occurred by the virtualization of agile practices. DaCamara et al. and Neumann et al. found, that the specific method of used agile practices is affected \cite{DaCamara.2020,Neumann.2021}. For example, they point to the use of tools like Retrium for retrospective meetings or the digitization of Kanban and Sprint Boards using Miro. Also specific techniques according to the effort estimation in planning meetings changed. Smite et al. investigated the effects of remote work on the agile practice pair programming \cite{Smite.2021}. The authors found that the use of pair programming decreased during Covid 19. They argue this with the increased effort of conducting the agile practice and a faster fatigue of the involved team members.

Another facet presented in several studies is the impact on social aspects, especially the communication and collaboration when using agile methods. Marek et al. \cite{Marek.2021} analyzed the impact of the switch to remote work on agile software development teams. The authors emphasize communication as an effect that has a positive impact on the work of agile software development teams through the stable productivity. In contrast, several authors describe a rather negative influence on communication and collaboration (e.g., \cite{Neumann.2021,Russo.2021a}). Neumann et al. \cite{Neumann.2021} referring to the challenge of intercollegiate communication and thus, a decreased social exchange between the team members. Various authors describe similar negative effects on the social aspects of agile software development teams (e.g., \cite{DaCamara.2020,Oconnor.2021}). Griffin describes that the risk of distractions during remote work is increased \cite{Griffin.2021}. 

To the best of our best knowledge, we found no peer reviewed studies in similar research context related to agile software development and the 4-day work week. Thus, we decided to search for literature dealing with the 4-day work week related to software development. Alfares presents a model for scheduling a 4-day work week, which aims to optimize the work organization and decrease the cost (and number) of employees \cite{Alfares.2003}. Also, we found grey literature related to our study. For instance, two white papers describe that a 4-day work week increases employee productivity as well as work motivation and satisfaction \cite{Andrews.2016,PerpetualGuardian.2019}. This results especially from the flexibility between the professional activity and the private environment. Furthermore, several articles discussing a experiment, which was performed by Microsoft (e.g., \cite{Chappel.2019,Eadicicco.2019}).

\section{Research Approach}
\label{S3_RDesign}
\subsection{Research Design}
We selected a case study approach and conducted the study based on the guidelines from Runeson and Hoest \cite{Runeson.2009}. We chose the exploratory research approach and argue our choice with the limited published research in the field. From our point of view, it is important to gain a deep understanding of how the agile software development teams react to the new situation working in a remote setting in a 4-day work week organization environment. Thus, we decided to select a qualitative research approach according to the guidelines from Yin \cite{Yin.2009}, Runeson and Hoest \cite{Runeson.2009}.

\begin{figure}[thb]
    \centering
	\includegraphics[clip,width=1\linewidth]{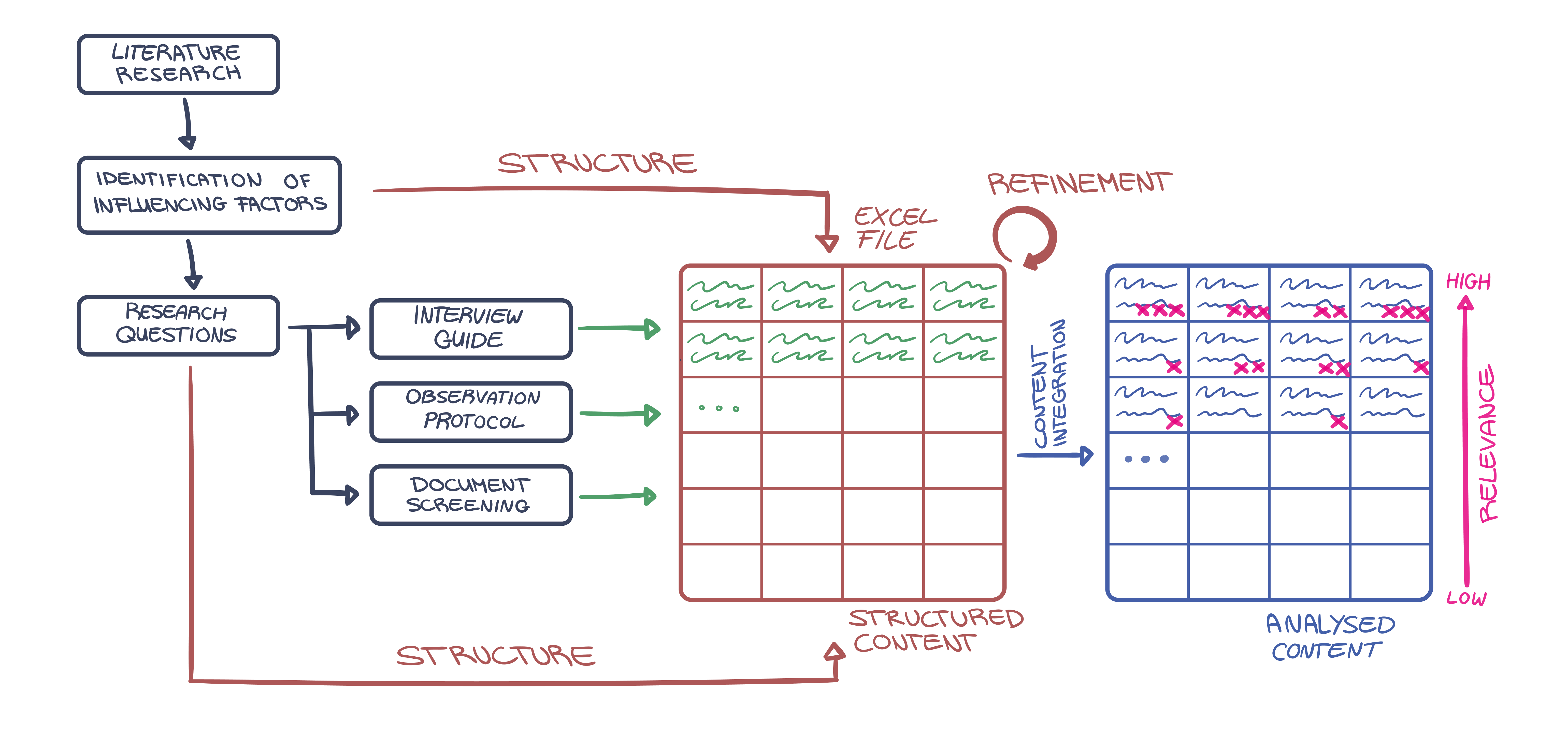}
	\caption{Research Design}
	\label{fig}
\end{figure}

Our research design is mainly organized in three steps. We present the research design in Figure \ref{fig}. First, we searched for existing literature in order to be able to identify the relevant influencing factors related to our topics 4-day work week and remote work during the Covid 19 pandemic. Based on these influencing factors we defined our research questions, which we present in the introduction. The research questions are the structural basis for our data collection methods. In a second step, we used the research questions to prepare the data collection, which we describe in detail in subsection 3.3. Based on the influencing factors and our research questions, we structured the data analysis (see subsection 3.4).

\subsection{Research Context}
We conducted our study at the company Pritchett Inc. (anonymized). The Pritchett Inc. is an online marketing company and operates worldwide. Approximately 1000 employees working for the company. 

As our study deals with agile software development, we focus on the software development departments of the company. Pritchett Inc. owns five software development departments in four countries (Germany, Poland, United Kingdom and United States of America). We performed the study in one software development department in Germany and conducted the data in two agile software development teams: Manny and Mitchell (both also anonymized).

Pritchett Inc. sent their employees worldwide to work from home caused by the Covid 19 pandemic in March 2020 and closed the offices for onsite work partially related to the Covid 19 situation in the country or region. The switch to remote work were new for the most of the employees at the German software development departments, as it was totally common to work onsite in the offices. However, the switch to remote work was supported by several tools, which were already in use by the teams. The agile software development teams under study used Microsoft Teams and Slack, also before the switch to remote work, because stakeholder and product owners are working from other departments, also from other countries. Actually, the company is organized by a work where ever, whenever you want principle. The employees can decide by themselves, if they want to work from home or be onsite in the office. This also applies to Pritchett Inc. departments in other cities or countries the employees want to go and work remote. 

In summer 2020, Pritchett Inc. started an experiment of a compressed work week. For the second half of 2020 the company switched to a 4,5-day work week. This experiment was adapted in January 2021, as Prittecht Inc. announced the switch to a 4-day work week. For the support of this major organizational change, the company provided several guidelines to the employees concerning aspects on vacation or illness. The employees have the opportunity to define one off day per week. The off day can vary from week to week. Also, the Pritchett Inc. management made clear, that the 4-day work week still has the status of an experiment. Today, this status is still active.

\subsection{Data Collection} 
As described above, we selected a qualitative research approach. We conducted the data in three ways in both agile software development teams: Semi-structured interviews with agile software development team members, observations of agile practices and team meetings and screening of documents from the software development teams and Prittchet Inc.

We conducted the semi-structured interviews in English based on a prepared interview guideline (see Appendix A). The interview guide consists of four phases: Information phase, warm-up and introduction phase, main phase and closing phase. In the information phase the interviewee get an introduction of the interviewer, a clarification of the objectives of the study based on prepared text phrases and organizational aspects, like asking the agreement of audio recording. The warm-up and introduction phase aims to collect specific information of the interviewee. For instance, what is the current role in the team. The main phase is organized based on the four research questions. In the closing phase we ask the interviewee for further questions or any aspects the person wants to add. The interview closes with thanks for participating. 

In total, we conducted seven interviews. An overview of the interviewees concerning their roles, teams and experiences is presented in Table \ref{Table1:ProfilesInterviewees}. Every interview was conducted by an interviewer and at least one other researcher, which protocols the interview. Also, we were able to record the audio of the interviews by the consent of the interviewees. Later, we created transcripts of each interview. We conducted all interviews with activated cameras on both sides: The interviewer and the interviewee. The interviews took an average of around 40 minutes. We conducted the interviews in English.

\begin{table}[h]
\centering
\begin{tabular}{p{1cm} | p{5cm} | p{2cm} | p{2cm}}
\hline
\textbf{ID} & \textbf{Current role} & \textbf{Team} & \textbf{Years of experience in ASD}\\
\hline
P01 & Scrum Master & Both teams & 12\\
\hline
P02 & Lead Engineer & Mitchell & 5\\
\hline
P03 & Lead Engineer & Manny & 13\\
\hline
P04 & Software Engineer & Mitchell & 1.5 \\
\hline
P05 & Software Engineer & Mitchell & 4\\
\hline
P06 & Software Developer & Mitchell & 12\\
\hline
P07 & Software Developer & Manny & 11.5\\
\hline
\end{tabular}
\caption{Profiles of the interviewees}
\label{Table1:ProfilesInterviewees}
\end{table}

The collected data through observation of agile practices and team meetings was documented in a standardized protocol (see Appendix B). We conducted 14 observations in total. An overview of the observations is given in Table \ref{Table2:Observations}. Every meeting was held virtually using Microsoft Teams. The observation was planned in collaboration with the Lead Engineers of the teams. Two researcher observed the agile practices meetings and documented their notes in the above mentioned protocol. After the observation the researchers cross-checked the collected data. The observed Sprint Plannings consists of the agile practices planning, retrospective and review meetings. The Sprint Planning 2 is used for the creation of work items related to the specific backlog items. 

\begin{table}[h]
\centering
\begin{tabular}{p{1cm} | p{5cm} | p{2cm} | p{2cm}}
\hline
\textbf{ID} & \textbf{Meeting/Agile Practice} & \textbf{Team} & \textbf{Nr. of participants}\\
\hline
B01 & Coffee Break & Both teams & 7\\
\hline
B02 & Sprint Planning & Mitchell & 9\\
\hline
B03 & Sprint Planning 2 & Mitchell & 4 \\
\hline
B04 & Sprint Planning & Manny & 10\\
\hline
B05 & Sprint Planning 2 & Manny & 5\\
\hline
B06 & Monthly department meeting & Both teams & 20\\
\hline
B07 & Daily Stand Up & Manny & 3\\
\hline
B08 & Daily Stand Up & Mitchell & 5\\
\hline
B09 & Coffee Break & Both teams & 5\\
\hline
B10 & Sprint Planning & Mitchell & 10\\
\hline
B11 & Sprint Planning 2 & Mitchell & 5\\
\hline
B12 & Sprint Planning & Manny & 8\\
\hline
B13 & Daily Stand Up & Manny & 10\\
\hline
B14 & Daily Stand Up & Mitchell & 10\\
\hline
\end{tabular}
\caption{Overview of the observations}
\label{Table2:Observations}
\end{table}

The third data source are several documents created by the agile software development teams and the Prittchet Inc. company. In total we screened eight documents. Three of theses documents are provided by Pritchett Inc.: Guideline for organization requirements related to the 4-day work week, guideline for requirements concerning the work wherever/when ever principle and an employee survey. The survey aims to gain an understanding of "drivers" (aspects) like the well-being, workload, management support or job satisfaction of the employees. 

The survey data was filtered to the department under study. The team related documents we screened are: Two team radar protocols from retrospective meetings, team internal guidelines concerning meeting organization and remote work. Finally, we checked the performance analysis data from the teams, which are exported from the task management system Jira. 

\subsection{Data Analysis} 
As shown in Figure \ref{fig}, our data analysis was done in three steps. First, we created an Excel file and used the research questions and identified influencing factors from the literature as a structural basis. In a second step we transferred our collected data to the structured Excel file. Based on the structured data in the Excel file we coded our data into 25 codes and eight categories. This coding was initially done close to our collected data  and refined by cross check iterations from the researchers. Third, we used the structured (coded and categorized) content to analyze, which information is more or less relevant for our study results. This was mainly done by content triangulation using a virtual whiteboard in Miro. We checked individually, which information can be found how often in the structured data content per data collection method and discussed our results in the researcher group. The more often information was identified, the relevance of the finding increases. Finally, the analyzed content provides us the possibility to evaluate the information ordered by their relevance according to the research questions. 

\section{Results}
\label{S4_Results}

\subsection{RQ 1: Effects on the agile method in use}
\label{S4-2_RQ1}
Due to the introduction of the 4-day work week, the agile method used has been adapted, because less time with a constant workload resulted in a tighter schedule. This adjustment was reflected in the statements made during the interviews. These revealed that the length of the sprints was reduced from two to one week (P01-P07), as P01 described: "\textit{We also did sprint time boxing and shortened our sprints from two weeks to one week.}". As a consequence of the shortened sprint length, agile practices related to the sprint change (sprint n \begin{math} \to \end{math} sprint n+1) were adapted. The affected agile practices are the planning, review and retrospective meetings. All of these agile practices were shortened in time (P01, P05). Several interviewees mentioned, that all non-urgent meetings were marked as optional as an additional adaption (P01, P02, P04, P06). For instance, P02 said: "\textit{In the past we had more spontaneous meetings that were not well prepared because we immediately go to a meeting room when a topic was coming up. Now when a topic is raising up we discuss if the meeting is needed.}". Non-urgent meetings are all those meetings that do not actively contribute to the productive progress of a project. In addition to the statements of the interviewees, the observations confirmed these results (B01-B14).

\subsection{RQ 2: Effects on the productivity} 
\label{S4-2_RQ2}
With regard to the productivity of the agile software development teams, the effects of remote work and the 4-day work week in relation to professional communication, effectiveness and stress were examined. We found that professional communication had become more efficient. All interviewees indicated that meetings are more coordinated and focused. Almost half of all interviewees declared that the number of meetings were decreasing and were taking less time (P01, P04, P07). The Scrum Master (P01) explains: "\textit{The number and the duration of meetings changed. So we also reduce the time for meetings as well where it was possible}". In addition, fewer private conversations and interruptions are taking place in meetings (P01-P04), as P03 describes: "\textit{The meetings in remote work are much more focused. There is less small talk and not that many interruptions during the meeting for example that someone needs a break or comes late. It’s easier to deal with meeting series}". The other interviewees noted that the number of meetings had not changed (P02, P05, P06). The different statements are probably affected due to the roles of the team members, as lead roles generally attend more meetings. Furthermore, discussions arising in meetings were overly technical (P01, P03, P04): "\textit{Communication is way more efficient communication and on point. In the office there were more small talks an personal conversations at the beginning and in the end of a meeting. Now we have nearly only work-related discussions in the online meetings.}"(P01). Nine out of 14 observed meetings showed that work-related communication was mostly not interrupted by private conversations (B03, B04, B06, B07, B08, B11, B12, B13, B14). Since the beginning of 2021, one project team  has set meeting guidelines in their Confluence space. They not only contain a code of conduct but also rules for meeting organization, participation, and documentation. This provides a better structured communication in meetings.  In both agile software development teams, staff absence days are tabulated in Confluence in relation to the 4-day week to ensure better coordination for meetings.

Similarly, we found that work had become more efficient since the switch to remote work. Six out of seven interviewees reported the same amount of work (P01, P02, P06, P07) or more work (P03, P04) in the same amount of time since the introduction of remote work. One Engineer describes: "\textit{Reducing one day per week, its obvious that this will produce overtime. But I have to mention that the company is still working 5 days a week, just the employees are working 4 days a week. So I just work further when I know I have my day off tomorrow.}" (P04). This was argued by the agile software development team members due to concentrated a focused manner while working from home because there are no decreased disruptive factors such as loudness, small talks with colleagues or other interruptions (P01-P07 and B01, B03-B09, B11, B12-B14): "\textit{Before we had a big office space where it was sometimes very loud and even if colleagues pass by we just have a short conversation what was kind of interrupting you.}" (P01). The working time on the four working days had basically increased (P01, P03-P06), as an engineer states (P06): "\textit{Yes, [I work more overtime] because we have our goals in the sprint planning. And if we see that the time is running out, we do some overtime to get these tasks finished.}". However, this could be attributed more to the project-related time pressure in the individual projects than to the 4-day work week or remote work (P05-P07). The statements from the interviews were also reflected in retrospective meetings. We verified retrospective protocols and found that since the deployment switch of remote work the workload increased temporarily but not continuously. In addition, the velocity report and the log of the solved tickets show a positive increase in the velocity and solved tickets in the time of the changeover to remote work and the 4-day work week. This strengthens our findings about the increase of efficiency.

\subsection{RQ 3: Effects on the job satisfaction and stress level}
\label{S4-2_RQ3}
\paragraph{Effects on the job satisfaction:}
The 4-day work week and remote work have a positive effect on the work of the agile software development teams by increasing job satisfaction: "\textit{Definitely [I like working from home]. I feel more productive. To go to office is more for socializing, team-building and workshops. Currently I think it’s the best way how we could work in the future.}" (P02). The quieter working place and flexible work scheduling enable a more productive work environment (P01-P03, P05-P07). Likewise, the elimination of commuting (P03-P06) and a better work-life balance (P06, P05) lead all respondents to be satisfied with remote work (P01-P07). The observations and document review confirm this by noting a calm work atmosphere without any interruptions in 11 out of 14 observed meetings (B01, B03-B14), as well as an upward trend in the "Satisfaction" section of the team radar since the introduction of remote work. 

In addition, the introduction of remote work leads to an increase in work motivation among the employees (P01-P05, P07). Only one respondent noted that his work motivation dependents on the project (P06), the engineer describes: "\textit{It always depends for me on the project, not on the remote work. For 4-day-work week the motivation is higher, because at the moment it’s a test phase so we need to be successful with that so we can continue. I think that’s what motivates everyone.}". The section "Accomplishment" in the employee survey reflects the statements of the interviewees. Since the introduction of remote work, satisfaction with work performance has increased. Besides, the sufficient provision of work equipment also contributes to employee satisfaction (P01-P07), as P03 explains: "\textit{We all have Notebooks from [Pritchett Inc.] and we are allowed to collect some hardware from the office.}". Due to defined guidelines in advance, the procurement of equipment for remote work is determined. These guidelines enable employees to obtain additional equipment at company's expense. This option was taken up by some respondents, for example, to get a better keyboard or a screen with a higher resolution (P02, P05). In addition to remote work, the 4-day week also leads to job satisfaction among employees because they feel happier, more balanced, and more satisfied (P01-P07). One lead engineer (P02) states: "\textit{I think the biggest change is that everyone is really happy with it. You see it in terms that everybody is motivated. Everyone seems very satisfied. The 4 day week is a real life changer.}". Reasons for this are the individually usable day off once a week, which offers more relaxation (P02, P04), the more flexible work schedule (P01-P06) and the additional time with the family (P01, P03). The Section "Engagement" and "Workload" in the employee survey reinforced these statements because firstly the employee satisfaction has steadily increased since the introduction of remote work and secondly a further increase is visible since the 4-day work week. Furthermore, employees do not perceive any negative impact on their workload, but rather draw positive effects from the introduction of remote work and the 4-day work week. 

\paragraph{Effects on the stress level of agile software development team members:} 
Due to the 4-day work week, the work-related stress of the employees had increased (P02, P04, P06, P07): "\textit{Sometimes it’s a bit more stressful than before. On some days the organization of tasks is harder because certain people may not be in office at that day because of 4-day work week. So, we need to organize more, that results in a bit more stress.}"(P06). According to the interviewees, this is related to the compressed workdays (P03, P06, P07) and the frequent context changes (P03, P07). However, for agile software development team members, an additional day off as well as the elimination of commuting time seem to be more important benefits (P01-P05, P07), as Engineer P04 states: "\textit{I would say because of one day more, its more flexibility and its relaxing my week more. I can plan this day as I want, so that’s improving my private life.}". Although work stress had increased, this did not lead to more intercollegiate conflicts (P02-P04, P06, P07) in the teams. This statement is also confirmed by the aspect "Peer Relationship" in the employee survey, in which the relationship between colleagues was examined. Here, there are no changes compared to the time before remote work. Despite the increased work-related stress, remote work and the 4-day work week leads to less stress in private everyday life due to more flexible leisure time and the elimination of commuting (P01-P06): "\textit{I am less stressed. Right now, I have no way to the office and back home. So, there I have no stress to get the train. I have a better work life balance right now and I am more flexible.}" (P05).

\subsection{RQ 4: Effects on the social culture} 
\label{S4-2_RQ4}
The 4-day work week has an impact on the social culture in the agile software development teams due to the compressed working schedule. This results mainly in a low willingness to participate in meetings with social context (B01, B09). Interviewees perceive these meetings as an interruption of active participation in the team (P01, P02, P05-P07). Engineer P07 states: "\textit{For these meetings, it’s always the same people who are participating in these kinds of meetings. Often, I don’t participate either because in Pair-Sessions we just keep working instead of taking part because we have no time for this.}". Some would replace them by the continued work as soon as there were time constraints (P01, P05, P07), the Scrum Master describes: "\textit{The acceptance of the personal online events is very rare because you skip these meetings instant when there is much work pressure due to the tight schedule. But some colleagues are taking these meetings every time. Some never come.}". The observations of meetings with a social focus confirm the interviewees' statements (B01, B09). Another effect on the social culture in the teams is caused by the more professional working environment mentioned by the interviewees (P02, P03, P05, P06). Reasons for this change were a propagated focus on work issues (P03) and a stronger separation of work environment and lunch break due to remote work (P06). In addition to the statements from the interviewees, the observed meetings also show a focused and goal-oriented execution. Furthermore, an efficient time management with adherence to time frames and no interruptions were observed (B03, B04, B06-B08, B11-B14). Besides, a low proportion of social communication in comparison to the total communication during these meetings was observed (B02, B04, B05, B08-B10). Despite the effects already mentioned, some interviewees assessed the relationship with their team members as unchanged (P02, P05-P07) or only slightly worse (P01, P02, P04). This estimation was also confirmed by internal documentation concerning the team member satisfaction, where no negative change in the relationships between team members can be seen.

\section{Threats to Validity}
\label{S5_ThreatstoValidity}
It is important to take several limitation into account when conducting case studies with a qualitative research method \cite{Runeson.2009}. 

\paragraph{Construct validity:} In this study, we considered the 4-day work week and remote work together, as both work organization types were used simultaneously at the time of the research. The design of our interview guideline counteract this aspect, as we designed the questions specifically to the two work organization types (remote work and 4-day work week). The interviews took an average of around 40 minutes. This length can be tiring for the interviewees and may lead to shorter answers towards the end of the interview than at the beginning. To counteract this effect, we conducted all interviews during regular working hours and pointed out that a time buffer of at least 10 minutes should be planned for the next scheduled appointment. Another aspect is the risk of not identifying all of the relevant literature, as we used recent studies to identify the influencing factors. Thus, we searched for related literature in several digital libraries and refined our search rings in iterative search runs. 

\paragraph{Internal validity:} Although a thorough analysis of recent literature was the basis for developing our interview guideline and observation protocol, some internal validity threats need to be taken into account. In order to avoid bias, we took several measures. First, the interview guideline consists of non-leading questions. Also, the interviews were designed as semi-structured. Thus, we were able to go in-depth in those directions the interviewee aims for. The interviewers did not personally know the interviewees. All interviews were conducted by at least two researchers. We also recorded every interview and created transcripts later. As a further measure, the researchers verified the transcripts from the recording, before we analyzed the data on detail. 

Furthermore, we used several triangulation types to strengthen the validity of our results as recommended by Runeson and Hoest \cite{Runeson.2009} and Yin \cite{Yin.2009}. We used different data collection types and sources. This triangulation helped us to optimize the consistency of our findings.  

\paragraph{External validity:} It is worth to mention, that the external validity could be higher with integrating more cases considered in Prittchet Inc. and in other companies, industries or countries. Further the phenomena under study should be affect other departments (e.g., marketing or human resources). Thus, a further analysis of non agile software development teams may be interesting, as the switch to the 4-day work week and remote work affects the other departments. 

\section{Conclusion and Future Work}
\label{S6_Conclusion}
This study presents our findings on the effects of a 4-day work week and the remote work of agile software development teams. In this section we summarize the results and provide ideas for future work.

In summary, the 4-day work week and remote work have various positive influences on the agile software development teams under study. First, the introduction of the two work organization models leads to an increased job satisfaction and productivity. However, we also found that the stress level of the team members increased. 

Second, the shortened work week and the resulting tighter schedule primarily affect the social exchange within the agile software development teams. In addition, the 4-day work week leads to adaptions of the agile method in use. Both the sprint length and several agile practices such as planning, review and retrospective meetings were adapted in particular concerning their length due to the compressed working time of the team members. Due to the compressing of the working time and the adaption of agile practices, the communication and execution of the agile practices is straight forward and become more formal. We confirm effects of the remote work presented by recent studies, such as the reduced social interaction among the team members. 

The 4-day work week and remote work seem to represent a flexible working model for the future to enable a better work-life balance and generally increase the job satisfaction and motivation of employees. To counteract the observed negative effects of the reduction of social communication, we recommend to implement regular workshops and events organized in onsite settings. 

In the context of this study, the 4-day work week and remote work were considered together, as both work organization models were used simultaneously at the time of our data collection. Future research could investigate which effects can be attributed to the 4-day work week or remote work in detail. This will gain a deeper understanding of these two work organization models and, where appropriate, provide new application scenarios and opportunities for organizing remote working agile software development teams. In addition, we recommend to transfer the research context to other settings to compare the two work organization models depending on aspects like the industry or company size.

\section*{Appendix A}
The interview guideline is available at the academic cloud: \href{https://sync.academiccloud.de/index.php/s/OlIg2XTuyQ4fcIZ}{Download Link}

\section*{Appendix B}
The observation protocol is available at the academic cloud: \href{https://sync.academiccloud.de/index.php/s/7WZDwRBNPjMnpCS}{Download Link}
%
%

\bibliographystyle{splncs04}
\bibliography{references}
\end{document}